\documentstyle[twocolumn,prb,aps,epsfig,floats,amssymb,html]{revtex} 

\newcommand{\figwidth}{0.9\columnwidth}

\begin{document}
\draft
\twocolumn[ \hsize\textwidth\columnwidth\hsize\csname@twocolumnfalse\endcsname

\title{Behavior of the thermopower in amorphous materials at the
  metal-insulator transition}

\author{C.\ Villagonzalo\cite{nip}, R.\ A.\ R\"{o}mer, and M.\ Schreiber}

\address{Institut f\"{u}r Physik, Technische Universit\"{a}t, 09107
  Chemnitz, Germany}

\author{A.\ MacKinnon}

\address{Blackett Laboratory, Imperial College, Prince Consort Rd.,
  London SW7 2BZ, U.K.}

\date{\today, $Revision: 1.14 $}
\preprint{SFB393/00-xx}

\maketitle

\begin{abstract}
  We study the behavior of the thermal transport properties in
  three-dimensional disordered systems close to the metal-insulator
  transition within linear response. Using a suitable form for the
  energy-dependent conductivity $\sigma$, we show that the
  value of the dynamical scaling exponent for noninteracting
  disordered systems such as the Anderson model of localization can be
  reproduced. Furthermore, the values of the thermopower $S$ have the
  right order of magnitude close to the transition as compared to the
  experimental results.
  A sign change in the thermoelectric power $S$ --- as is often
  observed in experiments --- can also be modeled within the linear
  response formulation using  modified experimental $\sigma$ data as input.
\end{abstract}


\narrowtext
] 

\section{Introduction}
\label{intro}

Transport phenomena in disordered quantum systems have been studied
for many years, \cite{KraM93,BelK94} yet many open problems remain. One
focus of these investigations is the metal-insulator transition (MIT).
This quantum phase transition from a good conducting material to an
insulator may be induced by disorder due to localization \cite{KraM93}
or by interactions such as electron-electron interactions and
electron-lattice coupling.\cite{BelK94,MotD79} In three-dimensional
(3D) amorphous materials the MIT is mainly attributed to
disorder.\cite{KraM93} For example, in heavily doped semiconductors
the disorder is brought about by the random distribution of dopant
atoms in the crystalline host.  However, indications of
electron-electron interactions have also been found, e.g., in the
d.c.\ conductivity $\sigma$ (or resistivity $\rho=1/\sigma$) in
doped semiconductors in both metallic\cite{RosATL81} and insulating
regimes.\cite{LiuSWZ96}

A further open problem is the behavior of the thermoelectric power $S$
or the Seebeck coefficient of disordered materials near the MIT. In
amorphous alloys and both compensated Si:(P,B) and uncompensated Si:P,
$S$ continuously changes from negative to positive values or vice
versa at low temperature $T$.  This corresponds to a change of thermal
conductors from electrons to holes or conversely and has been
attributed to electron-phonon interaction in amorphous alloys.
\cite{LauB95,SheHM91} On the other hand in heavily doped
semiconductors the sign change is believed to be caused by
electron-electron interactions or attributed to the existence of local
magnetic moments and their interactions with electrons.
\cite{LiuSWZ96,ZieLL96,LakL93} This conclusion is based on the
suppression of the anomalous behavior by a magnetic field.
\cite{ZieLL96,LakL93}
We remark that the sign change in $S$ is also observed in metals,
high-$T_c$ materials and quasicrystals.\cite{BerT96,ObeCT92,HabRKZ99}
Analytical treatments of metals as a degenerate free-electron gas
taking into account inelastic scattering with
phonons\cite{DurA94,DurA96,DurA00} as well as numerical considerations
incorporating electronic correlations in
superconductors\cite{HilHHG97} have also been shown to generate a sign
change in $S$. But in these systems the sign change occurs at a $T$
value which is 2 orders of magnitude higher than in disordered
systems.
Note that $S$ is mainly due to two distinct effects: (i) the diffusion
of the charge carriers and (ii) the net momentum transfer from phonons
to carriers.\cite{GalB92} But for $T<0.3$ K as considered in this
work, the diffusive part of the thermopower dominates that of the
phonon-drag contribution.\cite{GalB92,FlePRP00} Hence, from this point
on in this paper $S$ denotes only the diffusion thermopower.

The prototype for a theoretical description of 3D disordered systems
is the Anderson model of localization.\cite{And58} Near the MIT at $T=0$,
$\sigma$ behaves as
\begin{equation}
\sigma_{\text{c}} \propto \left\{
\begin{array}{ll}
  \sigma_0\left|1-\frac{E_{\text{F}}}{E_{\text{c}}}\right|^{\nu}, &
  |E_{\text{F}}| \leq E_{\text{c}}, \\ 
0, & |E_{\text{F}}| > E_{\text{c}},
\end{array}
\right. \label{sig:c}
\end{equation}
where $E_{\text{F}}$ is the Fermi energy, $E_{\text{c}}$ is the
mobility edge which separates the extended conducting states from
localized insulating states, and $\nu$ is a universal critical
exponent.  \cite{KraM93} By using Eq.\ (\ref{sig:c}) for $\sigma$ in a
linear response formulation the behavior of the thermoelectric
transport properties such as $S$,\cite{SivI86,KeaB88,EndB94,VilRS99a}
the thermal conductivity $K$ \cite{KeaB88,EndB94,VilRS99a}and the
Lorenz number $L_0$ \cite{EndB94,VilRS99a} at the MIT have been
computed.  Moreover, similar to $\sigma$, the quantities $S$, $K$
and $L_0$ have also been found to obey scaling.\cite{VilRS99b}
The scaling form of the dynamical conductivity $\sigma$ close to the MIT in 3D is
given as \cite{BelK94,LeeR85,AbrALR79,Weg76}
\begin{equation}
\label{scaling}
\frac{\sigma(t,T)}{T^{1/z}}={\mathcal{F}}\left(\frac{t}{T^{\nu z}}\right).
\end{equation}
Here $t$ measures a dimensionless distance from the critical point,
such as $t=(E_{\text{F}}-E_{\text{c}})/{E_{\text{c}}}$, the
correlation-length exponent $\nu$ in 3D is equivalent to the
conductivity exponent as given in Eq.\ (\ref{sig:c}), and $z$ is the
dynamical exponent.\cite{BelK94} For a noninteracting system such as
the Anderson model, one expects $z=d$ in $d$ dimensions.\cite{BelK94}
But, instead of obtaining $z=3$ in the scaling form of $\sigma$, one
finds $z\nu=1$. \cite{SivI86,VilRS99a,VilRS99b}
In addition to this discrepancy, $S$ turns out to be at least one
order of magnitude larger \cite{EndB94,VilRS99a} than the experimental
results in doped semiconductors \cite{LakL93} and in amorphous alloys.
\cite{LauB95,SheHM91} Furthermore, the sign change in $S$ cannot be
explained using the Anderson model and Eq.\ (\ref{sig:c}).
One may argue that the discrepancies between the transport
calculations and the experimental measurements are due to 
the absence of interactions in the Anderson model.  Indeed,
interactions may influence the behavior of the thermoelectric
transport properties. Yet we emphasize that the neglect of
interactions in the Anderson model is not entirely inconsistent with
the experimental situation in 3D amorphous materials. For example, recent
measurements in Si:P yield $\sigma$ scaling with $z\approx 3$ and
$\nu\approx 1$.\cite{WafPL99} This agrees with $z=d$ as predicted
by the scaling arguments \cite{BelK94,LeeR85} for noninteracting
systems.

The goal of this paper is to show that the correct value of $z$, the
right order of magnitude of $S$ at the MIT, and perhaps even the sign
change in $S$ at low $T$, can be described within a linear response
formulation using the noninteracting Anderson model of localization.
However, in order to do so, we have to use a more suitably chosen
energy-dependent $\sigma_{\text{c}}$ instead of Eq.\ (\ref{sig:c}).
After a brief review of linear transport theory, we construct a new
form for $\sigma_{\text{c}}$ as a function of energy $E$ and $T$ from
experimental data.  By using this model data as input for the linear
response formulation, we compute the temperature dependence of $S$,
$K$, $L_0$ and also $\sigma$ and show that they have the expected
qualitative and quantitative behavior close to the MIT. Finally, we
show that a small variation in $\sigma_{\text{c}}(E,T)$ can change the sign
of $S$. This effect cannot be produced simply by varying the density
of states $\varrho$ or the chemical potential $\mu(T)$.

\section{Linear Thermoelectric Transport Theory}
\label{linear}

In the presence of a small
temperature gradient $\nabla T$, the electric current density
$\langle{\mathbf{j}}_1\rangle$ and the heat current density
$\langle{\mathbf{j}}_2\rangle$ induced in a system are given (to
linear order) as
\begin{equation}
\langle{\mathbf{j}}_i\rangle=|e|^{-i}\left( |e|L_{i1}{\mathbf{E}}
-L_{i2}T^{-1}\nabla{T} \right), \label{current}
\end{equation}
where $e$ is the electron charge and $\mathbf{E}$ is the induced
electric field. $L_{ij}$ are the kinetic coefficients. Since we do
not consider the presence of a magnetic field in this work, the
Onsager relation $L_{ij}=L_{ji}$ holds.\cite{Cal85} Ohm's law, 
\begin{equation}
\langle{\mathbf{j}}_1\rangle=\sigma \mathbf{E},
\end{equation}
implies that in Eq.\ (\ref{current})
\begin{equation}
\sigma=L_{11}. \label{sigma1}
\end{equation}

The flow of thermal conductors due to $\nabla T$ is counteracted by an
electric force arising from $\mathbf{E}$ making
$\langle{\mathbf{j}}_1\rangle=0$. Equation (\ref{current}) then yields the
thermoelectric power $S$ which relates $\nabla T$ to $\mathbf{E}$,
\begin{equation}
S=\frac{L_{12}}{|e|TL_{11}}. \label{S1}
\end{equation}
The sign of $S$ determines whether the thermal carriers are electrons
or holes.  Using the Sommerfeld expansion for $|E_{\text{F}} -
E_{\text{c}}| > k_{\text{B}} T$, $S$ is given as
\cite{SivI86,KeaB88,VilRS99a,CasCGS88,AshM76}
\begin{equation}
  S=-\frac{\pi^2 k_{\text{B}}^2 T}{3|e|}\left.
    \frac{{\text{d}}\ln\sigma(E)}{{\text{d}}E}\right|_{E=E_{\text{F}}}\;,
\label{sommer}
\end{equation}
where $E_{\text{F}}$ is the Fermi energy, $k_{\text{B}}$ is Boltzmann's constant and
$\sigma(E)$ is assumed to be a slowly varying function on the scale of
$k_{\text{B}} T$.  Equation (\ref{sommer}) is also known as the Mott
formula.\cite{MotJ58}

The thermal conductivity $K$ determines the contribution to
$\langle{\mathbf{j}}_2\rangle$ stemming from $\nabla T$. Using Eqs.\ 
(\ref{sigma1}) and (\ref{S1}) in $\langle{\mathbf{j}}_2\rangle$ we
obtain $K$ in terms of the kinetic coefficient as
\cite{EndB94,VilRS99a}
\begin{equation}
K=\frac{L_{22}L_{11}-L_{21}L_{12}}{|e|^2TL_{11}}. \label{K1}
\end{equation}
From the definition of the Lorenz number it follows that
\cite{EndB94,VilRS99a}
\begin{equation}
L_0 \equiv \frac{e^2}{k_{\text{B}}^2}\frac{K}{\sigma T}
=\frac{L_{22}L_{11}-L_{21}L_{12}}{(k_{\text{B}} TL_{11})^2}. \label{L01}
\end{equation}
In metals at room $T$, $L_0=\pi^2/3$.\cite{AshM76}.  It also takes on
the same value at $T\lesssim10$ K in metals where the electrons suffer
no inelastic scattering processes.\cite{AshM76}

The primary consideration then in determining $\sigma$, $S$, $K$ and
$L_0$ is to calculate $L_{ij}$.  Under the assumptions that the system
is noninteracting and inelastic scattering processes are absent, 
$L_{ij}$ are given in the Chester-Thellung-Kubo-Greenwood formulation
\cite{CheT61,Kub57,Gre58} as
\begin{equation}
L_{ij}=\int^{\infty}_{-\infty} A(E)[E-\mu(T)]^{i+j-2}\left[
-\frac{\partial f(E,\mu,T)}{\partial E}\right]dE, \label{lij}
\end{equation}
where $i,j=1,2$, $\mu(T)$ is the chemical
potential, $f(E,\mu,T)$ is the Fermi distribution function and
$A(E)$ contains {\em all} the system-dependent features.

Lastly, we note that the $T$ dependence of $\mu$ can be obtained for
noninteracting systems from
\begin{equation}
n(\mu,T)=\int^{\infty}_{\infty}dE \varrho(E)f(E,\mu,T) \label{nmu}
\end{equation}
where $n$ is the number density of electrons and $\varrho$ is the density
of states.\cite{AshM76} Knowing $\varrho$ and keeping $n$ constant, we
find numerically that $\mu(T) \sim T^2$ in the 3D Anderson model
with an increased effective mass due to the disorder.\cite{VilRS99a}

\section{A Phenomenological Approach}
\label{phenomenology}

There are only two parameters that are model dependent in the
transport theory discussed in Sec.\ \ref{linear}. These are $A(E)$ and
$\mu(T)$. In order to determine the behavior of the thermoelectric
transport properties close to the Anderson MIT, 
$A(E)$ in Eq.\ (\ref{lij}) has usually been set
\cite{SivI86,KeaB88,EndB94,VilRS99a,CasCGS88} 
equal to the critical behavior of $\sigma$ given by Eq.\ 
(\ref{sig:c}). As mentioned in the introduction, this leads to the
unphysical value for $z=1/\nu$ and therefore an unphysical frequency
and $T$ dependence of $\sigma$. The main reason for this behavior is
easily understood: there is no $T$ dependence in Eq.\ (\ref{sig:c})
and thus all $T$ dependence in Eq.\ (\ref{sigma1}) is due to the
broadening of the Fermi function in Eq.\ (\ref{lij}) with increasing
$T$. Thus in order to model the correct $T$ dependence, we should add
to Eq.\ (\ref{sig:c}), valid at $T=0$, the desired $T$ dependencies
such as $\sigma\propto T^{1/z}$ in the metallic and $\sigma\propto
\exp{(-T)}$ in the insulating (say, variable-range-hopping) regimes.
Such a purely theoretical model for $\sigma_{\text{c}}(E,T)$ will then
incorporate a multitude of constants that can be adjusted to fit the
experimental results. Of course this is of limited practical use since
the validity of these fitting parameters is hard to justify.

Here we will instead use as input for $\sigma_{\text{c}}(E,T)$ recent
{\em experimental} data obtained by Waffenschmidt {\em et al}.
\cite{WafPL99} who measured $\sigma$ in Si:P at the MIT under uniaxial
stress. Their data yield good scaling of $\sigma$ according to Eq.\ 
(\ref{scaling}) with a dynamical exponent $z=2.94 \pm 0.3$ and $\nu =
1 \pm 0.1$. These values agree with the scaling arguments
\cite{BelK94,LeeR85} and reasonably well with the numerical results
\cite{SleO99,MilRS99a,MilRSU99} for noninteracting systems.

In the $\sigma(t,T)$ scaling of Ref.\ \onlinecite{WafPL99},
$t=(s-s_{\text{c}})/s_{\text{c}}$ where $s$ is the stress and
$s_{\text{c}}$ the corresponding value at the transition.  We sample
their scaled data for several values of $(t,T)$ and fit a spline curve
\cite{spline} $\sigma_{\text{c}}$ to these points in order to get a
smooth functional form for $\sigma(t,T)$. Transforming the spline
$\sigma_{\text{c}}$ as a function not only of $T$ but also of $E$, we
set $t=(E-E_{\text{c}})/E_{\text{c}}$.  Finally, we substitute
$\sigma_{\text{c}}(E,T)$ for $A(E)$ in Eq.\ (\ref{lij}).
Consequently, the thermoelectric transport properties are directly
obtained from Eqs.\ (\ref{sigma1})--(\ref{L01}).

In this paper we consider temperatures from $0.01$ K to $0.2$ K.  Far
from the transition we could not probe lower than $T<0.02$ K.  This is
due to the limited input data and consequently a limited range of the
spline function that generated $\sigma_{\text{c}}(E,T)$.  The unit of
$\sigma_{\text{c}}$ is taken as $\Omega^{-1}\mbox{cm}^{-1}$ consistent
with the experiments.  The $E$ scale is (arbitrarily) fixed at $1$ meV
which is the order of magnitude of the binding energy of an isolated
donor in a heavily doped semiconductor.  \cite{LiuSWZ96} In order to
compare with the previous results in the Anderson model
\cite{VilRS99a} we let $E_{\text{c}}=7.5$. We emphasize that this
value is of no significance and can be assigned (nearly) arbitrarily.
The important point to consider is the location of the Fermi energy
$E_{\text{F}}$ with respect to $E_{\text{c}}$.  This distinguishes the
electronic regimes.  Thus, the metallic, critical and insulating
regimes are identified as $|E_{\text{F}}|<E_{\text{c}}$,
$E_{\text{F}}=E_{\text{c}}$, and $|E_{\text{F}}|>E_{\text{c}}$,
respectively. Usually, $\mu(T)$ is derived in Eq.\ (\ref{nmu}) from
$\varrho$ of the 3D Anderson model of localization.  In the next section
we shall also show the effect of using a different functional form of
$\mu(T)$.

\section{Results and discussions}

\subsection{Temperature dependence of the thermoelectric transport properties}
\label{tempdep}

\begin{figure}
\epsfxsize=\figwidth \epsfbox{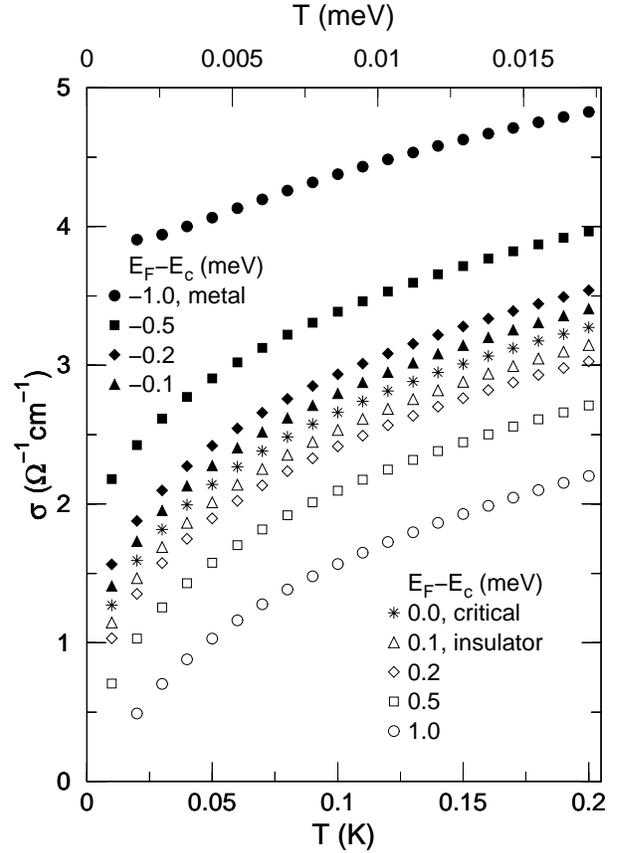}
\caption{
  Numerical calculations for the electrical
  conductivity, according to Eq.\ (\protect\ref{sigma1}), as a
  function of temperature $T$.  The filled symbols represent the
  metallic regime $|E_{\text{F}}|<E_{\text{c}}$, ($\ast$) denotes the
  critical regime $E_{\text{F}}=E_{\text{c}}$, and the open symbols
  represent the insulating regime $|E_{\text{F}}|>E_{\text{c}}$.}
\label{l11waffen}
\end{figure}

Consistent with the dynamics of the experiment in Ref.\ 
\onlinecite{WafPL99}, we expect $\sigma(T)\sim T^{1/z}$ at the
critical regime with $z\approx 3$. This is indeed the behavior of
$\sigma(T)$ close to $E_{\text{c}}$ as we show in Fig.\ 
\ref{l11waffen}.  For $|E_{\text{F}}-E_{\text{c}}|\leq 0.2$ meV we
obtain $z=3.2\pm0.3$. Note that $\sigma(T)=L_{11}$ has been integrated
according to Eq.\ (\ref{lij}) over the energy range where $\partial
f/\partial E \geq 10^{-20}$ meV.  Thus our numerical calculation of
$\sigma$ is consistent since it reproduces closely the original result
in Ref.\ \onlinecite{WafPL99}.  If we plot the results in Fig.\ 
\ref{l11waffen} with respect to
$(\mu-E_{\text{c}})/E_{\text{c}}T^{1/\nu z}$ we obtain a rough scaling
of $\sigma$ similar to Fig.\ 3 of Ref.\ \onlinecite{WafPL99}.

\begin{figure}
\epsfxsize=\figwidth \epsfbox{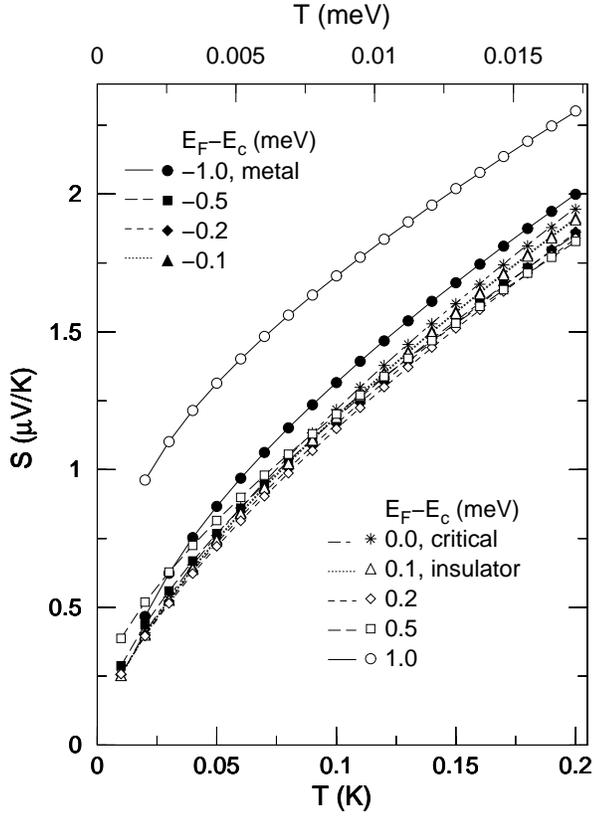}
\caption{
  The thermopower obtained from Eq.\ (\protect\ref{S1}) as a function
  of $T$ with the same symbols as in Fig.\ \protect\ref{l11waffen}
  distinguishing the metallic, critical and insulating regimes. The
  lines are obtained from Eq.\ (\protect\ref{sommer}). }
\label{swaffen}
\end{figure}

We next turn our attention to the thermoelectric power $S$. In the 3D
Anderson model of localization, we know that when using Eq.\ 
(\ref{sig:c}) one obtains $S\rightarrow 0$ in the metallic regime
\cite{SivI86,VilRS99a,CasCGS88} while in the insulating regime $S$
does not approach zero but seems to diverge as $T\rightarrow 0$.
\cite{VilRS99a,VilRS99b} At the MIT $S$ is a constant \cite{SivI86} of
the order of $100\;\mu$V/K.\cite{EndB94,VilRS99a,2dSiMOSFET} 
In Fig.\ \ref{swaffen}, we show that in the present approach $S$ in
the vicinity of the MIT is two orders of magnitude smaller compared to
previous results for the Anderson model.  The magnitude of $S$ is in
fact comparable to the experimental results in disordered
systems.\cite{LauB95,SheHM91,LakL93} Furthermore, $S\rightarrow 0$ as
$T\rightarrow 0$ in the metallic, critical and insulating cases.  This
behavior of $S(T)$ in all electronic regimes was observed
\cite{LauB95,Kar95} in amorphous Au$_x$Sb$_{1-x}$ and in amorphous
Ge$_{1-x}$Au$_x$. As indicated by the different lines in Fig.\ 
\ref{swaffen}, $S$ is in good agreement with Eq.\ (\ref{sommer}) since
$\sigma$ in Fig.\ \ref{l11waffen} is smooth across the transition at
finite $T$.
Note that in order to evaluate Eq.\ (\ref{sommer}) properly for the
system considered here, the $E$ dependence of the input spline
$\sigma_{\text{c}}(E,T)$ was used instead of $\sigma(T)$ from Fig.\ 
\ref{l11waffen}.
We emphasize that it is no contradiction that $S$ is positive here but
mainly negative in the doped semiconductors in all electronic
regimes.  In the energy regions close to $E_{\text{c}}>0$ the charge
carriers are holes instead of electrons as shown in Ref.\ 
\onlinecite{VilRS99a}. $S$ would be negative if we had chosen
the left mobility edge $E_{\text{c}}<0$ for low
filling.\cite{VilRS99a,VilR98}

\begin{figure}
\epsfxsize=\figwidth \epsfbox{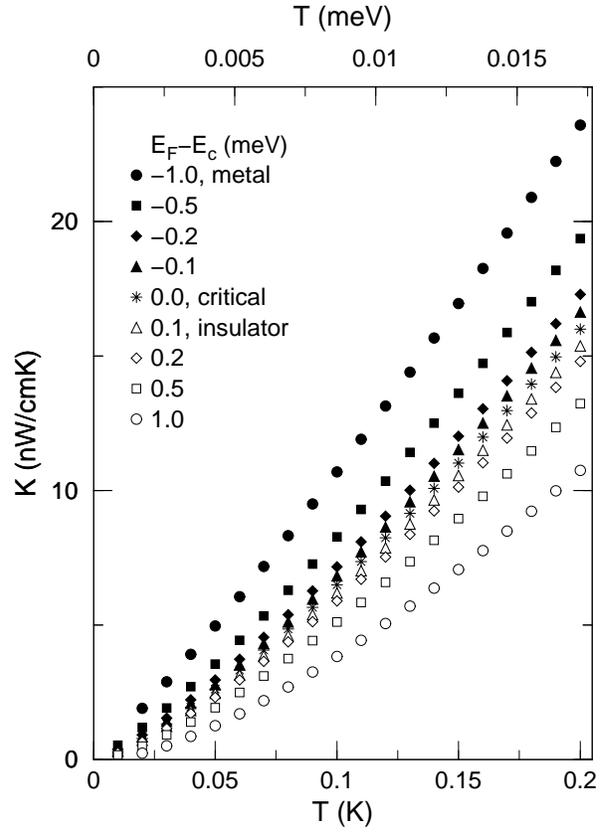}
\caption{
  The $T$ dependence of the thermal conductivity computed with Eq.\ 
  (\protect\ref{K1}).}
\label{kwaffen}
\end{figure}

\begin{figure}
  \epsfxsize=\figwidth \epsfbox{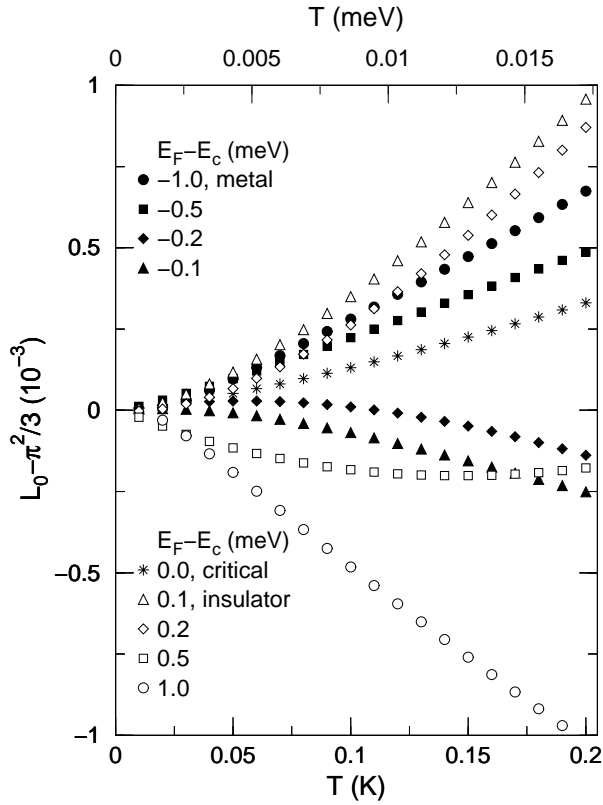}
\caption{
  The Lorenz number determined by Eq.\ (\protect\ref{L01}). The
  results are shifted by $\pi^2/3$, the universal value for
  metals.\cite{AshM76}}
\label{lowaffen}
\end{figure}

The corresponding $T$ dependence of $K$ is shown in Fig.\ 
\ref{kwaffen}. We find that $K\rightarrow 0$ as $T\rightarrow 0$ in
all electronic regimes. This is also the behavior of $K$ using
$\sigma_{\text{c}}$ in Eq.\ (\ref{sig:c}).\cite{VilRS99a} In the
metallic regime $K$ is larger than in the insulating regime since
there are more heat carriers in the former.  From the results of
$\sigma$ and $K$ in Figs.\ \ref{l11waffen} and \ref{kwaffen}
respectively, we obtain $L_0$.  As shown in Fig.\ \ref{lowaffen}, $L_0
\rightarrow \pi^2/3$ as $T\rightarrow 0$ whether it be in the
metallic, critical or insulating regime. This is different from the
results using Eq.\ (\ref{sig:c}) for $\sigma$.  There one obtains an
$L_0$ which depends on the conductivity exponent in the critical and
insulating regimes while it approaches the universal value
\cite{AshM76} $\pi^2/3$ only in the metallic regime.\cite{VilRS99a}
Here we see no markedly distinct behavior in the metallic regime
compared to the insulating regime. For
$|E_{\text{F}}-E_{\text{c}}|=0.1,0.2$ meV, $L_0$ in the metallic
regime is less than its corresponding value in the insulating regime.
For $|E_{\text{F}}-E_{\text{c}}|=0.5,1.0$ meV, $L_0$ in the metallic
regime is larger than its corresponding value in the insulating
regime.

In the calculation of $\sigma(T)$, $S(T)$, $K(T)$ and $L_0(T)$, we
used a phenomenological construction of $\sigma_{\text{c}}(E,T)$.
Furthermore, we have assumed that the density of states $\varrho$ is the
same as that of the 3D Anderson model of localization given in Ref.\ 
\onlinecite{VilRS99a}.  Since this $\varrho(E)$ is a smooth and (restricted to
$E>0$) monotonic function, $\mu(T)$ obtained from Eq.\ (\ref{nmu}) is
also smoothly and monotonically varying with $T$ as described in Sec.\ 
\ref{linear}.

\subsection{Effects of a structured $\varrho$}

\begin{figure}
\epsfxsize=\figwidth \epsfbox{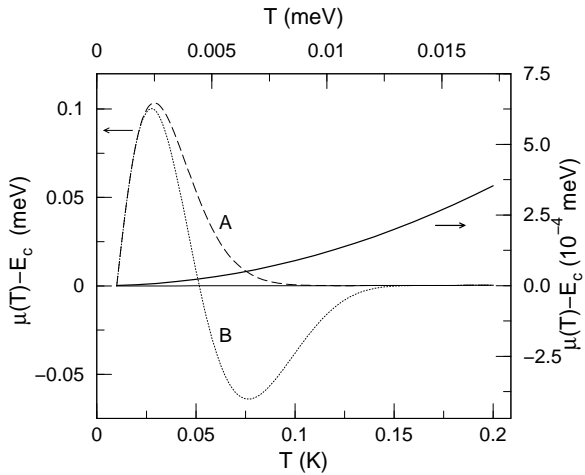}
\caption{
  Comparison between modified (dashed and dotted lines) and unmodified
  (solid lines) chemical potentials. The curves are shifted with
  respect to the mobility edge. The thick solid line is the unmodified
  $\mu$ shown on a finer (right) scale.}
\label{mubump}
\end{figure}

We now consider the effects of a possible structure in $\varrho$.  We
shall assume here that this structure corresponds only to variations
in $\mu(T)$ and not in $\sigma$.  In Fig.\ \ref{mubump} we show two
examples of a modified $\mu(T)$ in the critical regime.  Example A has
a pronounced maximum, while example B has both a maximum and a
minimum.  The height of the maximum in both examples A and B is
$\approx 0.1$ meV.  Note that this is significantly larger than the
halfwidth of the bump which is $\leq 0.005$ meV.  This is also true
for the depth of the minimum in example B. Thus a small change in $T$
corresponds to a large change in $\mu(T)$.  Applying these forms of
$\mu(T)$ together with Eq.\ (\ref{sig:c}) for $\sigma$ reproduces the
same structures in $S$.  For example, using form B of $\mu(T)$ we
obtain an $S$ having both a maximum and a minimum in the same $T$
interval as $\mu(T)$.  But $S$ is still of the order of $100\mu$V/K
while the variations are only of the order of $10\mu$V/K and not
large enough to cause a sign change in $S$.  On the other hand using
the phenomenological construction of $\sigma_{\text{c}}(E,T)$ yields
even smaller changes. We observe variations in $L_{11}$ and $L_{12}$
of less than 10$\%$ from their unmodified values.  Consequently, we
find negligible changes in $S$.  Figure \ref{swaffen} would appear
unmodified.  Hence we conclude that even a large change in the density
of states $\varrho$ and thus also in $\mu(T)$ is not sufficient to cause
the change of sign for $S$ as observed in the experiments. However,
this weak dependence of $S$ on $\varrho$ and $\mu(T)$ at least justifies
our use of the simple Anderson density of states in the present paper.

\subsection{Effects of a structured $\sigma(E)$}

\begin{figure}
\epsfxsize=\figwidth \epsfbox{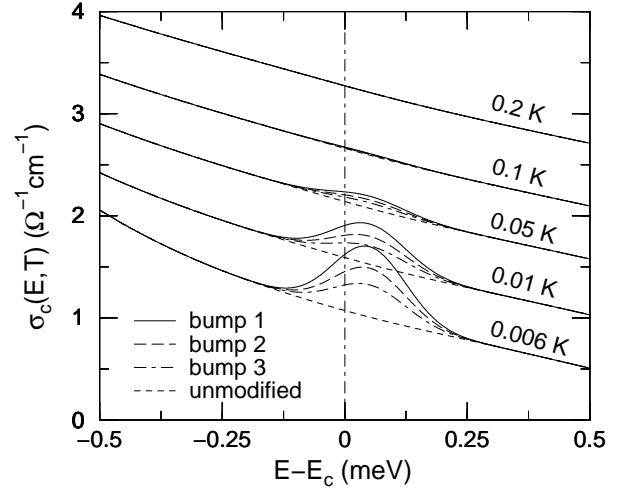}
\caption{
  The modified $\sigma_{\text{c}}(E,T)$ as input having increasing bumps with
  decreasing $T$ centered near at $E-E_{\text{c}}=0.05$ meV.  For clarity only
  selected isotherms are shown. The vertical line indicates the
  mobility edge.}
\label{dcbump}
\end{figure}

Let us now assume that for small $T$ there are non-monotonicities in
$\sigma_{\text{c}}(E,T)$ --- although these have not been observed in the
experiments.\cite{LauB95,SheHM91,LakL93} Thus we consider the case
when there is a sizeable change in $\sigma_{\text{c}}(E,T)$ in the region
close to $E_{\text{c}}$ for small $T$.  The corresponding `bumps' in
$\sigma_{\text{c}}(E,T)$ are shown in Fig.\ \ref{dcbump} with different peak
heights and with half-widths $<1$ meV. For simplicity they are
essentially quadratic functions of $E$ and have been generated such
that they decay quickly as $\exp(-T^4)$ with increasing $T$. The height of the
bumps is $<1\;\Omega^{-1}$cm$^{-1}$ which is at least an order of
magnitude smaller than the values observed for $\sigma$ in the
measurements.\cite{LauB95,WafPL99} The lowest temperature studied is
$T=6$ mK and we shall only consider metallic and insulating regions
with $|E_{\text{F}}-E_{\text{c}}|\leq 0.1$ meV.

\begin{figure}
\epsfxsize=\figwidth \epsfbox{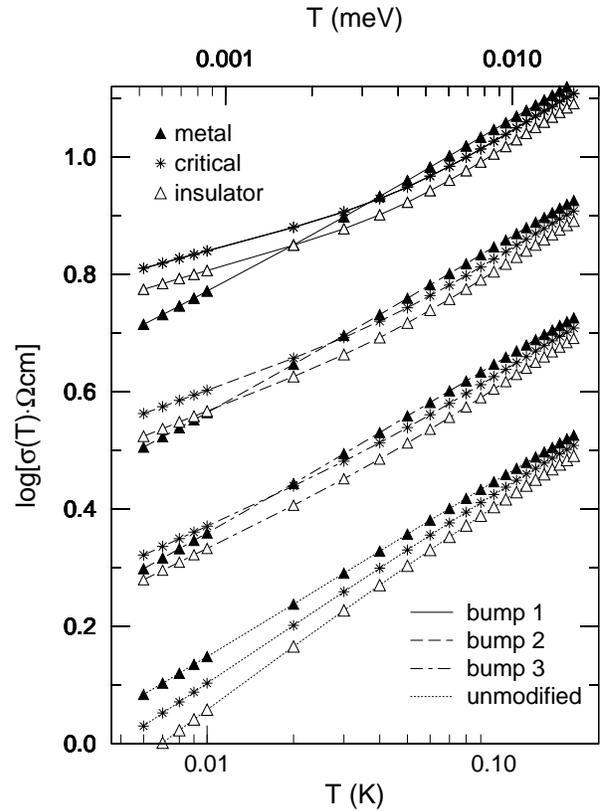}
\caption{
  Comparison between $\sigma$ with and without bumps in the metallic,
  critical and insulating regimes for
  $E_{\text{F}}-E_{\text{c}}=-0.1,\;0.0,\;0.1$ meV.  For clarity each
  set of $\sigma(T)$ is shifted by $0.2$ along the vertical axis from
  each preceding set. The lines are guides for the eye only.}
\label{l11bump}
\end{figure}

Our results in Fig.\ \ref{l11bump} using $\sigma_{\text{c}}(E,T)$ with
and without bumps indicate that there are only small variations in the
slope of $\log(\sigma)$ and $\sigma\sim T^{1/3}$ remains valid within
the accuracy of these estimations.  We remark that the lowest measured
temperature in Ref.\ \onlinecite{WafPL99} is $15$ mK.  From Fig.\ 
\ref{l11bump} we see that the variations for $T\geq 15$ mK are much
smaller than for $T< 15$ mK.  Hence, these variations in $\sigma$
could not have been observed in the experiments.

\begin{figure}
\epsfxsize=\figwidth \epsfbox{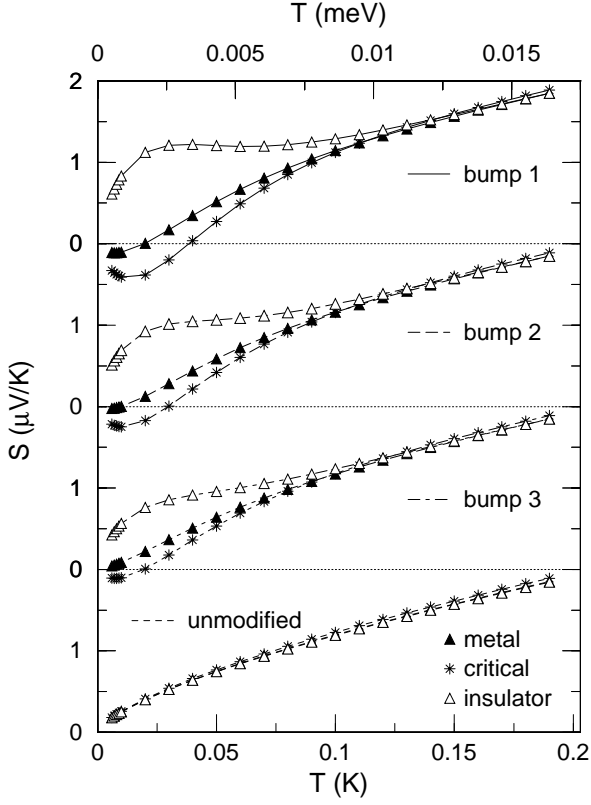}
\caption{
  The thermopower for different cases of $\sigma_{\text{c}}$ input.
  The lines are guides for the eye only.}
\label{sbump}
\end{figure}

In Fig.\ \ref{sbump} we show how the bumps affect $S$.  Even with the
very small bump 3, $S$ changes sign in the critical regime as
$T\rightarrow 0$.  As the bump increases this change becomes more
pronounced.  The temperature $T_{S=0}$ at which the sign change occurs
is $0.1$ K for bump 3, $T_{S=0}=0.2$ K for bump 2, and $T_{S=0}\approx
0.4$ K for bump 1.  These results for $T_{S=0}$ are still about one
order of magnitude less than in semiconductors \cite{ZieLL96,LakL93}
and two orders of magnitude smaller than in amorphous
alloys.\cite{LauB95,SheHM91} Of course, as shown in Fig.\ \ref{sbump},
$T_{S=0}$ shifts to higher values as the bump height increases.
Nevertheless, the minimum value of $S$ for $T<T_{S=0}$ has the same
order of magnitude as the corresponding maximum value of $S$ in
Si:(P,B) and in Si:P. We emphasize that the value of $T_{S=0}$  of
course depends on the energy unit chosen and thus will vary for
systems with different bandwidths, e.g., a larger band width will give
rise to a larger value of $T_{S=0}$.
We observe a similar sign change in the metallic regime but the depth
of the minimum is smaller than in the critical regime. The Mott
formula (\ref{sommer}) with $\sigma(T)$ given in Fig.\ \ref{l11bump}
can readily model this behavior since $\sigma(T)$ remains slowly
varying even if $\sigma_{\text{c}}(E,T)$ has a bump near
$E_{\text{c}}$.  In the insulating regime, $S$ has a shallow maximum
and drops back to zero as $T\rightarrow 0$.  This is different from
experiment,\cite{LiuSWZ96} where $S$ changes sign and neither has a
maximum nor minimum in the insulating regime.

\begin{figure}
\epsfxsize=\figwidth \epsfbox{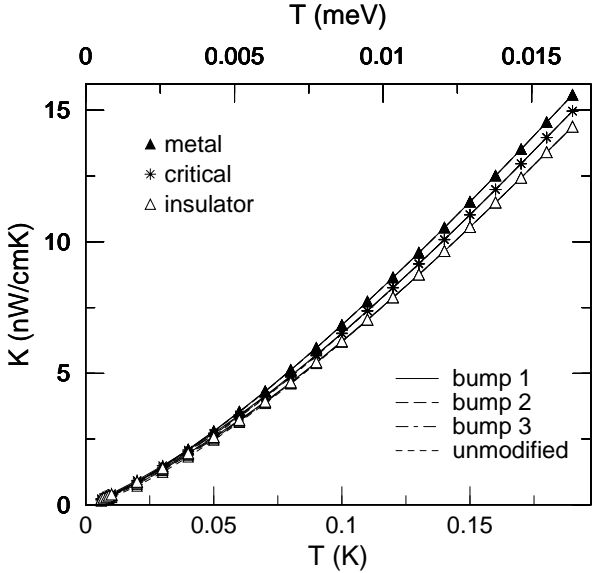}
\caption{
  The thermal conductivty remains largely unaffected by variations in
  $\sigma_c(E,T)$. The data for unmodified and modified
  $\sigma_c(E,T)$ lie on top of each other. The lines are guides for
  the eye only.  }
\label{kbump}
\end{figure}

\begin{figure}
\epsfxsize=\figwidth \epsfbox{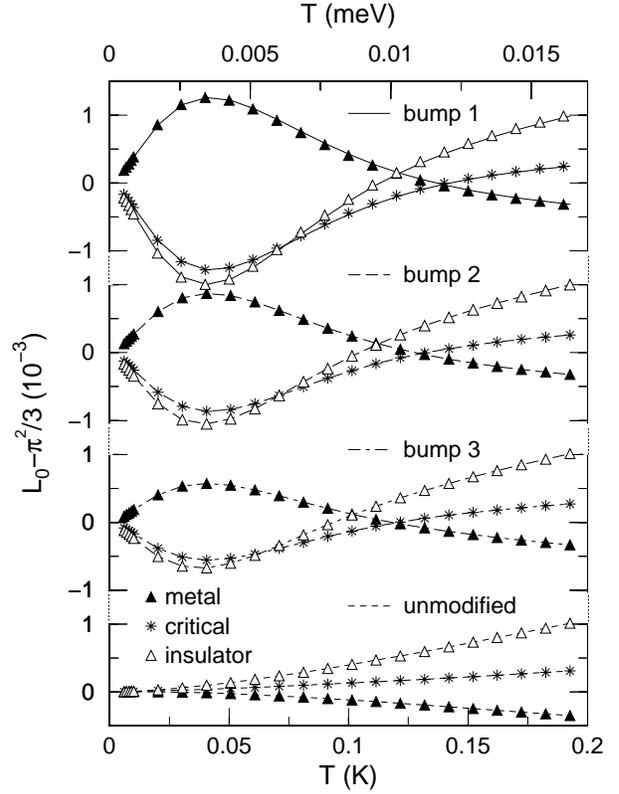}
\caption{
  The variation of the Lorenz number with $T$ shifted by $\pi^2/3$ for
  different cases of $\sigma_{\text{c}}$ input.  The lines are guides
  for the eye only.  }
\label{lobump}
\end{figure}

Unlike in $S$, there is no dramatic change in $K$ as can be seen in
Fig.\ \ref{kbump}. We find only negligible variations at $T<15$ mK.
This should be expected since there has also been hardly any change in
the slope of $\sigma(T)$ except for $T<15$ mK.
However, the small increase in $\sigma$ at $T<15$ mK in the metallic
regime together with the minimally modified $K$ leads to a drastic
change in $L_0$ even in the case for the smallest bump. The increase and
decrease in $\sigma$ leads to a maximum and minimum in $L_0$,
respectively.  However, $L_0$ still approaches the universal value
$\pi^2/3$ for $T\rightarrow 0$ as demonstrated in Fig.\ \ref{lobump}.

\section{Conclusions}

In this paper, we have shown that the anticipated value of the
dynamical scaling exponent $z\approx 3$ as well as the right order of
magnitude for the thermopower $S\approx 1 \mu$V/K at the MIT can be
obtained when taking into account the expected $T$ dependencies in
addition to the simple scaling behavior of Eq.\ (\ref{scaling}). Our
approach is phenomenological in the sense that we have refrained from
using fitting parameters and have rather taken experimental data as
input.  Using this data, we can explain the large deviations from
experimental results as reported in the theoretical studies of Refs.\ 
\onlinecite{SivI86,KeaB88,EndB94,VilRS99a,VilRS99b,CasCGS88}. 
We have shown that our results for $S$ agree with those predicted by the
Mott formula (\ref{sommer}) since we have used a $\sigma$ slowly varying
on the scale of $k_{\text{B}} T$ near the MIT. 
We emphasize, however, that for a disordered system where interactions 
are negligible, we should still expect the Anderson-type transition
as given in Eq.\ (\ref{sig:c}) at $T=0$.
Consequently, $S\approx 100\mu$V/K at the MIT
\cite{SivI86,KeaB88,EndB94,VilRS99a} should again be expected and one
should observe a large increase of $|S|$ at very low $T$.  However,
such temperatures appear presently inaccessible by experiment.

As a further challenge, we considered the sign changes observed in $S$
at low $T$. We found that even large variations in the chemical
potential $\mu(T)$ do not lead to a sign change in $S$. On
the other hand, a variation in the input $\sigma_{\text{c}}(E,T)$ data can give
rise to such a sign change in $S$, while at the same time resulting in
only small changes in the conductivity $\sigma$. 
Hence we have effectively modeled the underlying physical reasons for
the sign change --- which have been attributed to electron-electron
interactions or to the existence of local magnetic moments and their
interactions with electrons\cite{LiuSWZ96,ZieLL96,LakL93} or to
inelastic scattering with phonons\cite{DurA94,DurA96,DurA00} --- by simply
changing the input $\sigma_{\text{c}}(E,T)$.
Regarding a possible test for the existence of such a structured
$\sigma_{\text{c}}(E,T)$, we have shown that the $T$ variation of
$L_0$ is much more sensitive to the bumps than $\sigma$. Thus we have
been able to describe the main features of the critical behavior of
$S(T)$ although it remains unclear what might cause bumps in
$\sigma_{\text{c}}(E,T)$ close to $E_{\text{c}}$.  A microscopic and
possible system-dependent approach to the problem may eventually
account for these abrupt changes in $\sigma_{\text{c}}$.  Of course,
if many-particle interactions and electron-phonon coupling are
important, we no longer expect the feasibility of the
Chester-Thellung-Kubo-Greenwood formulation \cite{CheT61,Kub57,Gre58}
used here.

\section*{Acknowledgment}

This work was supported by the DFG through Sonderforschungsbereich
393, by the DAAD, the British Council and the SMWK. The authors
would like to thank R.\ Fletcher, R.\ Rentsch, R.\ Rosenbaum and B.\ 
Sandow for useful discussions. We thank R. Rosenbaum for a critical
reading of the manuscript.


\begin{thebibliography}{99}
\bibitem{2dSiMOSFET}
  This behavior of $S$ as $T\rightarrow 0$ is very similar to that
  observed in 2D electron gases in high-mobility Si-MOSFET's in what
  appears to be a MIT.\cite{FlePRP00} However, it is not clear why $S$
  in these materials behaves like that of a $3d$ Anderson model.  In
  addition, $S$ at the claimed MIT has an order of magnitude of $\sim
  10\;\mu$V/K.\cite{FlePRP00}
\end{thebibliography}



\end{document}